\documentclass{elsart}
\usepackage{amssymb}
\usepackage{graphicx}

\newcommand{\figurewidth}{10cm}
\begin{document}
\runauthor{Binder, Luijten, M\"uller, Wilding, Bl\"ote}
\begin{frontmatter}
\title{Monte Carlo investigations of phase transitions: status and
perspectives}

\author[Mainz]{Kurt Binder}
\author[Mainz]{Erik Luijten\thanksref{mpi}}
\author[Mainz]{Marcus M\"uller}
\author[Mainz]{Nigel B. Wilding\thanksref{Edin}}
\author[Delft]{Henk W.J. Bl\"ote}

\address[Mainz]{Institut f\"ur Physik, Johannes Gutenberg-Universit\"at,
Staudinger Weg 7, D-55099 Mainz, Germany.}
\address[Delft]{Department of Physics, Delft University of Technology, P.O. Box
5046, 2600 GA Delft, The Netherlands.}
\thanks[mpi]{Also at Max-Planck-Institut f\"ur Polymerforschung, Postfach
3148, 55021 Mainz, Germany. E-mail: erik.luijten@uni-mainz.de}
\thanks[Edin]{Present address: Department of Physics and Astronomy, The
University of Edinburgh, Edinburgh EH9 3JZ, U.K.}

\begin{abstract}
  Using the concept of finite-size scaling, Monte Carlo calculations of various
  models have become a very useful tool for the study of critical phenomena,
  with the system linear dimension as a variable. As an example, several recent
  studies of Ising models are discussed, as well as the extension to models of
  polymer mixtures and solutions. It is shown that using appropriate cluster
  algorithms, even the scaling functions describing the crossover from the
  Ising universality class to the mean-field behavior with increasing
  interaction range can be described. Additionally, the issue of finite-size
  scaling in Ising models above the marginal dimension ($d^*=4$) is discussed.
\end{abstract}

\end{frontmatter}


\section{Introduction}

It is a common belief that at the present time, about 30 years after the
renormalization-group theory of critical phenomena was invented~\cite{1},
static critical behavior of systems in thermal equilibrium is rather well
understood. In particular, this is expected to be true for the most intensively
studied case, the Ising universality class~\cite{2}, to which systems such as
uniaxial ferromagnets, binary alloys, simple fluids, fluid mixtures, polymer
solutions and polymer blends belong~\cite{3}. However, in the present work, we
shall draw attention to some aspects of critical behavior in Ising-like spin
systems which are, even today, still incompletely understood. The first of
these concerns the problem of crossover between the Ising universality class
and mean-field critical behavior. This crossover occurs, for instance, when the
interaction range (and hence the ``Ginzburg number''~$G$ entering the Ginzburg
criterion~\cite{4}) is varied~\cite{5,6,7,8,9,10,11}. A closely-related
crossover is found for symmetrical polymer mixtures when the chain length $N$
of the polymers is varied~\cite{3,12,13,14,15,16,17,18,19,20,21}. A part of
this crossover (though typically not the full extent of the crossover scaling
function) can be probed experimentally near the critical point of fluids and
fluid binary mixtures~\cite{22,23,24}. While the Ginzburg
criteria~\cite{4,13,14} provide a qualitative understanding of this crossover,
the quantitatively accurate theoretical prediction of the crossover scaling
function is a challenging problem~\cite{25,26,27,28,29,30,31,32,33}, and hence
Monte Carlo studies~\cite{5,6,7,8,9,10,11,15,16,17} are of great potential
benefit. In particular, the question as to what extent (if at all) such
crossover scaling functions are universal is an intriguing
one~\cite{9,10,11,22,23,33}.

Another very interesting crossover which can also be studied is that which
occurs near the critical point of unmixing for polymer solutions in a bad
solvent~\cite{12,34,35,36,37,38,39,40,41,42,43}. For chain length $N\to\infty$
the critical temperature $T_{\rm c}(N)$ moves towards the $\Theta$-temperature,
where a single coil undergoes a transition from a swollen coil to a collapsed
globule. This limit corresponds to a tricritical point~\cite{12}.

Monte Carlo analyses of critical phenomena typically apply finite-size scaling
concepts~\cite{44,45,46,47,48,49}. However, care is necessary in the proper
application of these methods in the mean-field limit. In fact, the standard
formulation of finite-size scaling (``linear dimensions $L$ scale with the
correlation length $\xi$'') implies that the hyperscaling relation~\cite{2}
between critical exponents should hold \cite{45,50}, which is not the case for
mean-field exponents (apart from $d=d^*=4$ dimensions). This problem already
arises for Ising models with short-range interactions for
$d>d^*$~\cite{51,52,53,54,55,56,57,58,59,60,61}, and some disagreements between
Monte Carlo results~\cite{51,52,54} and theoretical predictions~\cite{53,60}
have stimulated a long-standing debate (see~\cite{61} for a detailed review).

\section{Mean-field to Ising crossover}

We consider the Hamiltonian~\cite{5,6}
\begin{equation}
{\mathcal H}/k_{\rm B}T = - \sum_i \sum_{j > i} K({\bf r}_i-{\bf r}_j) s_i s_j
                      - h_0 \sum_i s_i \;,
\label{eq:latham}
\end{equation} 
with $s_i = \pm 1$ and an interaction $K({\bf r}) \equiv cR^{-d}$ for $|{\bf
r}| \leq R$ and zero elsewhere. The critical behavior of this model on
$d$-dimensional lattices can be studied efficiently with a new cluster
algorithm adapted for long-range interactions~\cite{62}.

To analyze the crossover it is instructive to consider the associated
Ginzburg--Landau field theory in continuous space,
\begin{eqnarray}
  \mathcal{H}(\phi)/k_{\rm B}T &=& - \int_{V} d{\bf r}  \left\{ \frac{1}{2}
  \int_{|{\bf r}-{\bf r}'| \leq R} d{\bf r}' \left[ \frac{c}{R^d} \phi({\bf r})
  \phi({\bf r}') \right] - \frac{1}{2} v \phi^2({\bf r}) \right.\nonumber \\
  && \left. \phantom{-\int_{V} d{\bf r}  \left\{
            \vphantom{\int_{|{\bf r}-{\bf r}'|}\left[\frac{c}{R^d}\right]}
            \right.}
  - u_0 \phi^4({\bf r}) + h_0\phi({\bf r} \right\} \;,
\label{eq:hamil}
\end{eqnarray}
where $\phi({\bf r})$ is the single-component order-parameter field, $v$ is a
temperature-like parameter and $u_0$ is a constant. After Fourier
transformation and suitable rescaling this can be rewritten as (here $N$ is the
total number of lattice sites)
\begin{eqnarray}
  \bar{{\mathcal H}}/k_{\rm B}T &=& \frac{1}{2} \sum_{{\bf k}} \left[
  k^2 + \frac{r_0}{R^2} \right]
  \psi_{\bf k} \psi_{-{\bf k}} \nonumber \\
  && + \frac{u}{4 R^4 N}
  \sum_{{\bf k}_1} \sum_{{\bf k}_2} \sum_{{\bf k}_3}
  \psi_{{\bf k}_1} \psi_{{\bf k}_2} \psi_{{\bf k}_3}
  \psi_{-{\bf k}_1-{\bf k}_2-{\bf k}_3} -
  \frac{h}{R} \sqrt{\frac{N}{2}} \psi_{{\bf k}={\bf 0}} \;,
\label{eq:scal-hamil}
\end{eqnarray}
where $u$ is related to~$u_0$ and $h$ to $h_0$~\cite{6}, and $r_0$ in
mean-field theory is the deviation of the temperature from its critical-point
value.

\begin{figure}
\centerline{\includegraphics[width=\figurewidth]{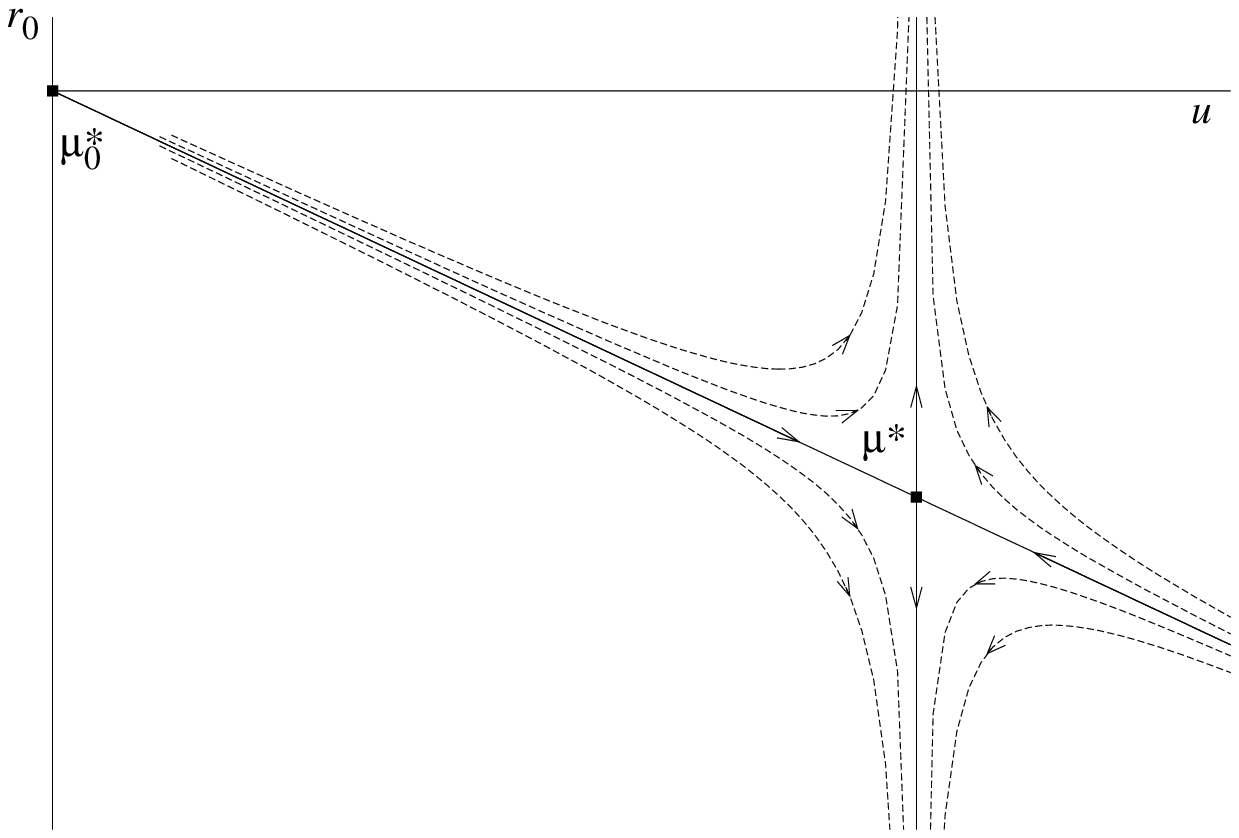}}
\caption{Qualitative picture of the renormalization trajectory describing the
crossover from the Gaussian fixed point $\mu_0^*=(0,0)$ to the Ising fixed
point $\mu^*=(r_0^*,u^*)$.}
\label{fig:1}
\end{figure}

We are now interested in identifying the crossover scaling variable associated
with the crossover from the Gaussian fixed point $u=0$ and $r_0=0$ to the
nontrivial Ising fixed point (Fig.~\ref{fig:1}). Because of the trivial
character of the Gaussian fixed point and the fact the crossover scaling
description should hold all the way from the Ising fixed point to the Gaussian
fixed point, one can infer the crossover length scale $l_0 = R^{4/(4-d)}$
exactly! This is done by considering a renormalization by a length scale~$l$,
such that the wavenumber changes from~$k$ to $k'=kl$, the number of degrees of
freedom is reduced from $N$ to $N'=Nl^{-d}$, and $\psi_k$ changes into
$\psi'_{k'} = l^{-1}\psi_k$ to leave $\bar{{\mathcal H}}$ invariant. From
inspection of the terms in the Hamiltonian one can conclude that the singular
part of the free energy must satisfy the scaling relation
\begin{equation}
\tilde{f}_{\rm s}\left( \frac{r_0}{R^2}, \frac{u}{R^4}, \frac{h}{R} \right)
 = l^{-d}\tilde{f}_{\rm s}\left(\frac{r_0}{R^2}l^2,
  \frac{u}{R^4}l^{4-d}, \frac{h}{R}l^{1+d/2}\right)  \;.
\label{eq:enerscal1}
\end{equation}
We see that a finite and nonzero value for the second argument of
$\tilde{f}_{\rm s}$ is retained exactly when $l$ takes the value of the
crossover scale~$l_0$. Thus we conclude that the singular part of the free
energy scales with~$R$ as follows
\begin{equation}
\tilde{f}_{\rm s} = R^{-4d/(4-d)} \hat{f}_{\rm s} \left(\tilde{r}_0
  R^{2d/(4-d)}, \tilde{u}, h R^{3d/(4-d)} \right) \;,
\label{eq:enerscal}
\end{equation}
where a natural choice of coordinates (Fig.~\ref{fig:1}) is to measure
$\tilde{r}_0$ and $\tilde{u}_0$ as distances from the Ising fixed point, unlike
in the original Hamiltonian, where $r_0$ and $u_0$ are distances from the
Gaussian fixed point.

Equation~(\ref{eq:enerscal}) describes how the temperature distance
$\tilde{r}_0$ from criticality and the magnetic field~$h$ scale with the range
of interaction~$R$: Note that the crossover exponent is known exactly (unlike
other cases of crossover, e.g., between the Ising and Heisenberg universality
class in isotropic magnets with varying uniaxial anisotropy~\cite{63}). The
same result for the crossover exponent follows~\cite{5}, of course, from
simple-minded arguments using the Ginzburg criterion. However, the location of
the nontrivial fixed point~$u^*$ (Fig.~\ref{fig:1}), the associated other
exponents, and the explicit form of the scaling function~$\tilde{f}_{\rm s}$
cannot be obtained exactly.

\begin{figure}
\centerline{\includegraphics[width=\figurewidth]{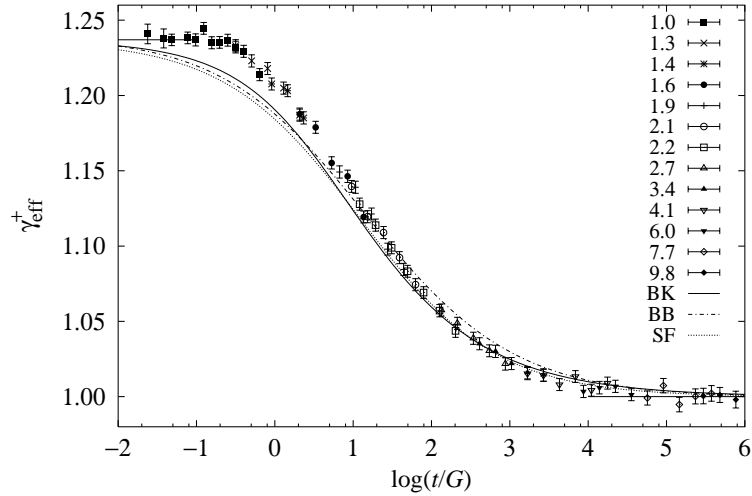}}
\caption{Effective susceptibility exponent $\gamma^*_{\rm eff}$ above $T_{\rm
c}$ for the three-dimensional Ising model with variable interaction range $R$
(numbers in the key) plotted vs.\ $t/G$, along with three theoretical
calculations for this quantity; due to Refs.~\protect\cite{31}~(BK),
\protect\cite{27}~(BB), and~\protect\cite{25,29}~(SF), respectively. From
Ref.~\protect\cite{9}.}
\label{fig:2}
\end{figure}

The calculation of the scaling function for the free-energy density or its
derivatives, such as the susceptibility, is a nontrivial task for both
renormalization-group and Monte Carlo calculations. This is demonstrated in
Fig.~\ref{fig:2} where the effective critical exponent $\gamma^{+}_{\rm eff}$
of the susceptibility for $T>T_{\rm c}$ is plotted versus the thermal crossover
scaling variable~$t/G$, with $t=(T-T_{\rm c})/T_{\rm c}$ being the reduced
temperature and $G=G_0 R^{-6}$ the Ginzburg number in $d=3$, for which $G_0
\approx 0.277$. Note that effective exponents are defined as
\begin{equation}
 \gamma^{\pm}_{\rm eff} \equiv -d \ln \hat{\chi} / d \ln |t| \;, \quad
 \hat{\chi} \equiv k_{\rm B}T_{\rm c}(R) (\partial M /\partial h)_{T} \;,
\label{eq:gamma_eff}
\end{equation}
where $\pm$ refers to $T \gtrless T_{\rm c}$, respectively, $M = \langle s
\rangle_{T, h}$, and the range~$R$ is defined from the second moment of the
interaction ($z$ being the effective coordination number)
\begin{eqnarray}
R^2 &=& \sum_{j \neq i} |{\bf r}_i - {\bf r}_j|^2 K({\bf r}_i - {\bf r}_j) /
        \sum_{j \neq i} K({\bf r}_i - {\bf r}_j) \nonumber \\
    &=& \frac{1}{z} \sum_{j \neq i} |{\bf r}_i - {\bf r}_j |^2
        \quad \mbox{with } |{\bf r}_i - {\bf r}_j | \leq R_m \;.
\end{eqnarray}
Here the second equality holds only for a square-well potential and values
$R_m^2 = 1, 2, 3, 4, 5, 6, 8, 12, 18, 28, 60, 100$, and~$160$ were studied.
From Fig.~\ref{fig:2} we see that the Monte Carlo results agree with all the
theoretical calculations near the Gaussian fixed point, but do not yield the
more rapid increase of $\gamma^{+}_{\rm eff}$ near the Ising fixed point.  It
is not clear what precise conclusions should be drawn from this discrepancy:
All these theoretical treatments really rely on extrapolations of low-order
renormalization-group expansions in $\varepsilon = 4-d$, and hence are perhaps
rather inaccurate in $d=3$ dimensions. On the other hand, they clearly relate
to the limit where $R \to \infty$ and $t \to 0$, with $tR^6$ fixed---a
universal description of the crossover can only be expected in this limit. The
Monte Carlo data shown in Fig.~\ref{fig:2} also include the range of small~$R$,
for which additional corrections to scaling present near the Ising fixed point
(other than those attributable to the Ising--mean-field crossover) may come
into play.

A rather successful description of the Monte Carlo data could be obtained by a
fit to a function given by Anisimov {\em et al.}~\cite{33}. Their description
is also an interpolation formula based on low-order $\varepsilon$-expansions
but contains a second parameter (in addition to~$G$) describing a
short-wavelength cutoff. However, one disturbing feature of this description is
that one needs different amplitudes $G_0$ in the relation $G = G_0 R^{-6}$
above and below~$T_{\rm c}$, and the ratio $G_0^{+}/G_0^{-}$ is an additional,
ad hoc, parameter the significance of which is not understood~\cite{33}.  Thus
we consider it an as yet unsettled problem as to just on which parameters the
crossover scaling description should depend.  In this context, we draw
attention to the question whether the specific square-well form chosen for the
exchange interaction matters. To answer this question, a more general form of
$K({\bf r}_i-{\bf r}_j)$ was chosen (viz., a superposition of two square-well
potentials which differ in range and strength but are chosen such that the same
value for~$R^2$ results~\cite{10}. While $T_{\rm c}$ was shown not to be
determined by~$R$ alone, but depended on $K({\bf r}_i-{\bf r}_j)$ in a more
detailed way, the same crossover scaling function resulted for all choices of
the interaction profile studied~\cite{10}.

\begin{figure}
\centerline{\includegraphics[width=\figurewidth]{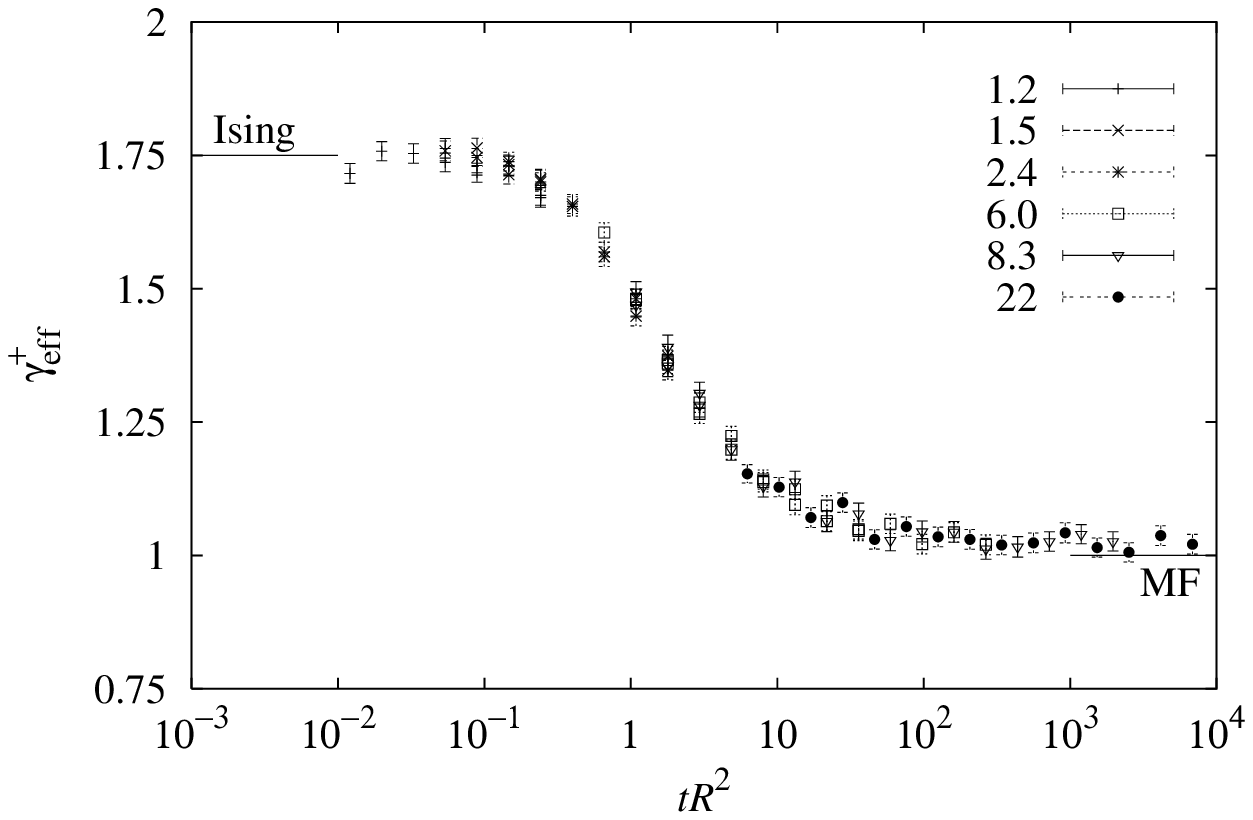}}
\centerline{\includegraphics[width=\figurewidth]{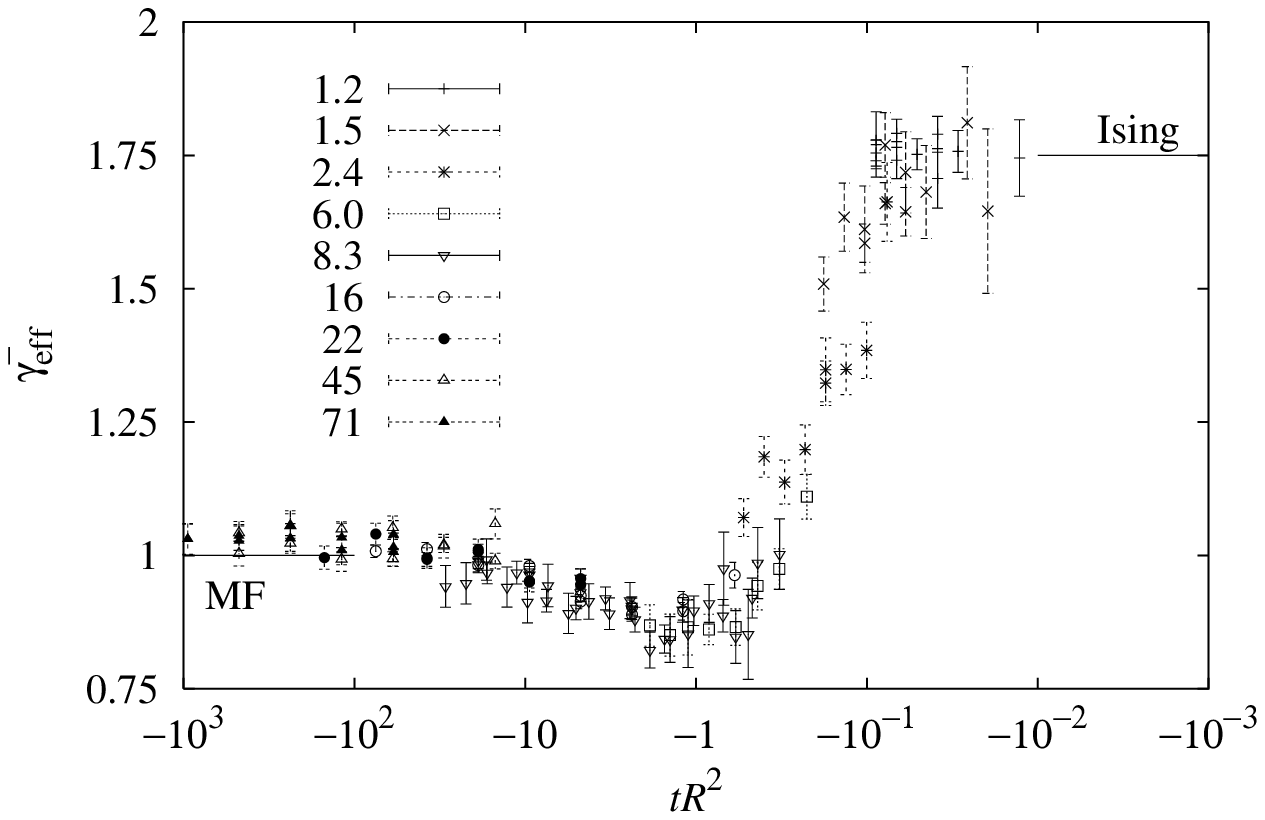}}
\caption{The effective susceptibility exponent $\gamma^*_{\rm eff}$ above
$T_{\rm c}$ (a) and below $T_{\rm c}$ (b), for the two-dimensional Ising model
with variable interaction range $R$ (numbers in the key), plotted vs.\ $tR^2$.
From Ref.~\protect\cite{8}.}
\label{fig:3}
\end{figure}

A particular merit of the description of Anisimov {\em et al.}~\cite{33} is,
however, that it can yield a non-monotonic variation of $\gamma^{-}_{\rm eff}$
with $t/G$: In $d=3$ a shallow minimum ($\gamma^{-}_{\rm eff} \approx 0.96 <
\gamma_{\rm MF} = 1$) occurs for $|t|R^6 = 10^2$~\cite{9} that can be fitted by
this theory~\cite{33}. Indeed, a very similar minimum has been observed in
Ref.~\cite{pelissetto98} from a mean-field expansion for Ising systems with
medium-range interactions, see also Ref.~\cite{pelissetto99} for a detailed
review. In $d=2$ dimensions, such a minimum occurs as well and is much more
pronounced than in $d=3$, while above $T_{\rm c}$ the variation of the
effective exponent is still monotonic (Fig.~\ref{fig:3}). Note that the
crossover is again spread out over many decades in the crossover variable~$t/G$
($G \propto R^{-2}$ in $d=2$), as in $d=3$, and that for $T<T_{\rm c}$ there
are no analytical results whatsoever to compare our Monte Carlo results with!
At this point, there is clearly still a gap in our knowledge about critical
phenomena.

As a last point in this section, we add a few brief comments about the way in
which the Monte Carlo results on effective exponents have been obtained. As
is well known~\cite{45,46,47,48,49}, the Monte Carlo method converges to the
exact statistical mechanics of a finite system only; the thermodynamic limit is
never addressed directly. The typical situation is that one deals with a
$L\times L$ or $L\times L\times L$ box with periodic boundary conditions.
The critical singularities are rounded and shifted by the finite size of the
system~\cite{44,45,46,47,48,49}. For the precise location of the
critical point, a finite-size scaling analysis is required. The principle of
finite-size scaling is that the linear dimension $L$ scales with the
correlation length $\xi$. Therefore the $k$'th moment of the magnetization $m$
scales like:
\begin{equation}
\left\langle |M|^k \right\rangle = L^{-k\beta/\nu} \tilde{M}_k(L/\xi) \;,
\label{eqn:8}
\end{equation}
$\beta$ and $\nu$ being the critical exponents of the order parameter ($\langle
m \rangle \propto |t|^\beta$) and the correlation length ($\xi \propto
|t|^{-\nu}$), respectively, and $\tilde{M}_k$ being some scaling function.
Therefore the straightforward observation results~\cite{45} that these power
law prefactors $L^{-k\beta/\nu}$ cancel out if one considers suitable ratios of
moments, such as
\begin{equation}
Q = \langle M^2\rangle^2/\langle M^4\rangle=\tilde{Q}(L/\xi) \;.
\label{eq:Q}
\end{equation}
At $T_{\rm c}$ we have $\xi \to \infty$, of course, so $\tilde{Q}(0)$ is simply
a constant, independent of the system size $L$. This justifies the
simple recipe to record this ratio for different choices of $L$ and obtain
$T_{\rm c}$ from the intersection point of these ratios~\cite{45,48,49}. Note
that the ordinate value of this intersection point is a universal constant
(only depending on the shape of the system and on the boundary conditions, but
not on $R$, for instance, provided one is in the asymptotic critical region).

However, this recipe so far ignores the crossover from one universality class
to the other (as well as corrections to scaling).  Nevertheles, it turns out
that one can formulate a combined finite-size scaling and crossover scaling
description for such problems~\cite{5,6,7,8,9,10,11,16,17,64}.  A simplified
description considers the variation of the correlation length, which is $\xi
\propto Rt^{-1/2}$ in the mean-field critical region, and $\xi \propto
(R^\kappa t)^{-\nu}$ in the Ising critical region [the exponent $\kappa$
follows from the condition that for $t=t_{\rm cross} \propto R^{-2d/(4-d)}$ and
the corresponding value of $\xi$, $\xi_{\rm cross}=\xi(t=t_{\rm cross})=l_0
\propto R^{4/(4-d)}$ a smooth crossover between both power laws occurs]. Now it
is of crucial importance to compare $L$ with the crossover length scale
$\xi_{\rm cross}$: If $L$ is much less than $\xi_{\rm cross}$, then the finite
size rounding occurs fully in the mean-field regime, before the crossover to
Ising criticality has had a chance to come into play. Actually in this regime
the correlation length $\xi$ is not the relevant length to describe the finite
size rounding~\cite{48,51,52}, one rather needs the so-called ``thermodynamic
length''~\cite{52}, $\ell_T \propto |t|^{-2/d}$, as will be discussed in
Sec.~4.  In this regime ($L \ll \xi_{\rm cross}$) an accurate determination of
$T_{\rm c}$ is clearly impossible. In order to accurately locate $T_{\rm c}$,
we need to study the inverse regime, $L \gg \xi_{\rm cross}$: Only then can one
see the mean-field critical behavior farther away from $T_{\rm c}$ crossing
over to the Ising behavior at $t_{\rm cross}$ (remember that this crossover is
spread out over several decades!) and the finite-size rounding sets in at a
still much smaller value of $|t|$ (where $L\simeq\xi$). Since for large $R$,
$\xi_{\rm cross}$ is also very large ($\xi_{\rm cross}\simeq l_0\propto
R^{4/(4-d)}$), one needs to simulate very large $L$ and hence such simulations
are technically very difficult.

\begin{figure}
\centerline{\includegraphics[width=\figurewidth]{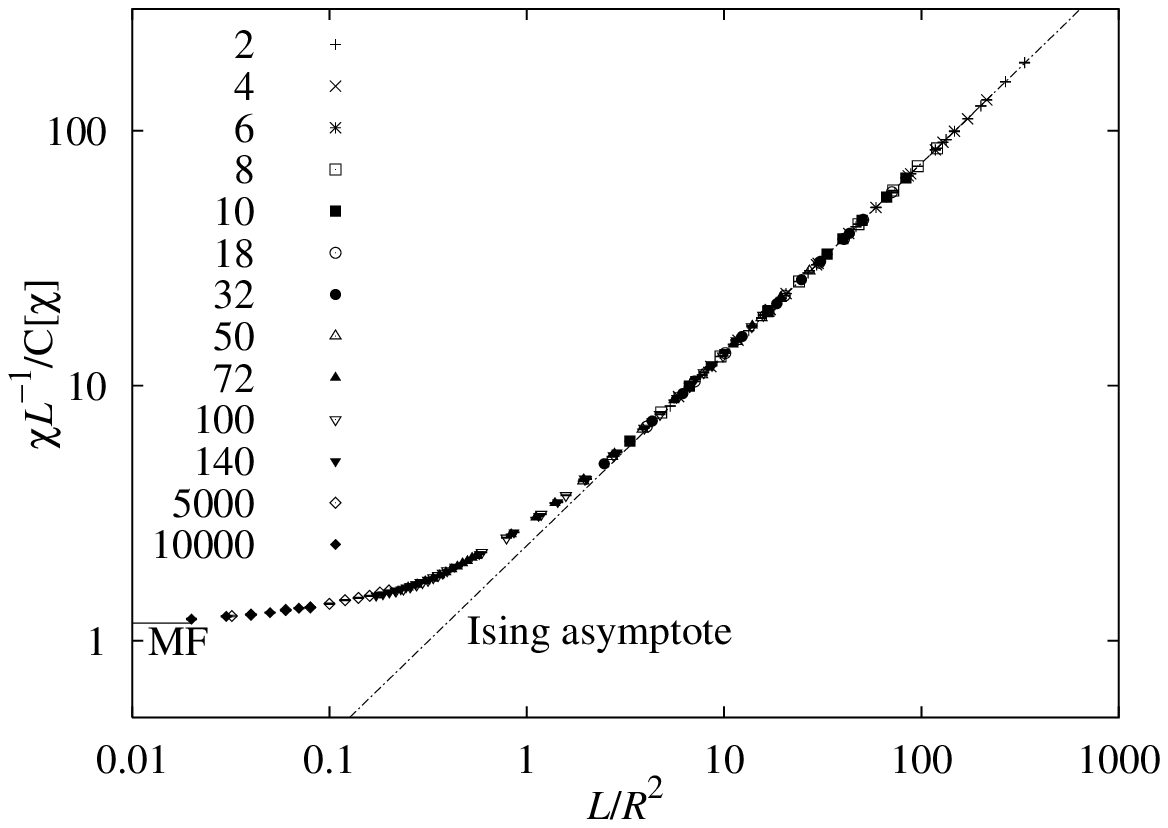}}
\centerline{\includegraphics[width=\figurewidth]{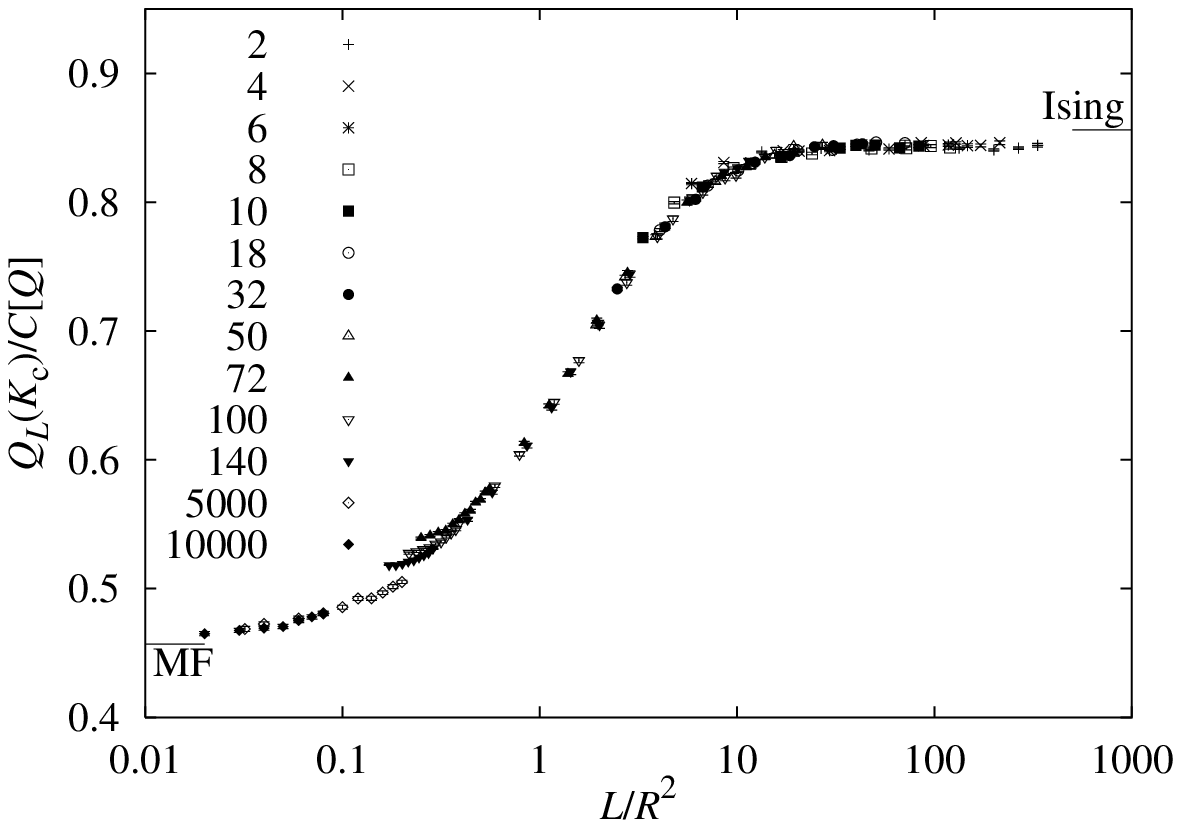}}
\caption{Finite-size crossover curve for the magnetic susceptibility $\chi$
divided by the system linear dimension (a) and the amplitude ratio $Q$
[Eq.~(\protect\ref{eq:Q})] (b) for the two-dimensional Ising model at
$K=K_c(R)$ plotted vs.\ the finite-size crossover scaling variable $L/R^2$
(note that $\xi_{\rm cross}=l_0 \propto R^2$ in $d=2$).  In both quantities,
range-dependent correction factors $C[\chi]$ and $C[Q]$ have been divided out
to eliminate some corrections to scaling (see Ref.~\protect\cite{7} for a
definition of these factors). From Ref.~\protect\cite{7}.}
\label{fig:4}
\end{figure}

Thus it is not surprising that when this problem was first addressed with
single-spin-flip Monte Carlo algorithms~\cite{5} a satisfactory description of
the full crossover could not be obtained, and the availability of an efficient
cluster algorithm~\cite{62} was crucial for obtaining meaningful results. In
$d=2$, we could study $L$ up to 800 lattice units, and $R_m=100$ corresponding
to $z=436$ interacting neighbors (Fig.~\ref{fig:4}). With these large lattices
it is possible to follow the variation of $Q$ almost all the way from the
mean-field limit at small $L$ to the Ising limit at large $L$, and in $\chi$
(Fig.~\ref{fig:4}(a) the Ising asymptote (slope $3/4$ on the log--log plot) is
nicely confirmed).

Since we know $T_{\rm c}$ very precisely and have data for such a wide range of
$L$, it is also possible to carry out runs slightly away from $T_{\rm c}$,
which are used to study the thermal crossover presented in
Figs.~\ref{fig:2},~\ref{fig:3}. Only data not affected by the finite system
size are used for the numerical derivative required in
Eq.~(\ref{eq:gamma_eff}).

\section{First steps towards the study of crossover problems in polymer blends
and solutions.}

As is well known~\cite{3,12,13,14}, the Ginzburg--Landau--Wilson Hamiltonian
for a symmetrical polymer mixture near its critical unmixing point can be
mapped on to the Ising model with a medium range of interaction (in $d=3$
dimensions), $N^{1/2}$ (with $N$ being the chain length of the flexible
macromolecule) playing the role of the interaction volume $R^3$. Qualitatively,
this mapping is understood from the fact that a polymer coil has a random
walk-like configuration. Its gyration radius $R_{gyr}$ scales as $R_{gyr}\simeq
a\sqrt{N/6}$, where $a$ is the size of the monomer. Thus the monomer density
of one chain inside the volume that is occupies ($V\propto R^3_{gyr}$) is very
small, $\rho=N/V\propto a^{-3}N^{-1/2}$. Hence in a dense melt
($\rho_{\rm melt}\simeq a^{-3}$) there are $N^{1/2}$ chains in the same volume,
i.e.\ each chain interacts with $x=N^{1/2}$ ``neighbors''. Thus as
$N\to\infty$ one again expects a crossover from Ising-like critical behavior to
mean-field like behavior, and this is verified experimentally~\cite{18,19}
(though the corresponding prediction for the Ginzburg number $G\propto 1/N$
does not seem to work out.

First steps to study this crossover by computer simulation have been performed
\cite{16,17} using the bond fluctuation model of symmetrical polymer
mixtures~\cite{9,65} applying a semi-grand canonical algorithm~\cite{15} and
histogram reweighting techniques~\cite{66}. The model and methodology of these
simulations have been extensively reviewed elsewhere~\cite{3,65} and hence we
omit all the technical details here, and simply show an attempt to estimate the
crossover scaling function of the order parameter~\cite{17} (Fig.~\ref{fig:5}).
Note that polymer are slowly relaxing objects and hence difficult to
simulate---no counterpart to the cluster algorithm used for the Ising
model~\cite{62} is available, and hence the challenge remains to improve
substantially the accuracy of studies such as shown in Fig.~\ref{fig:5} in
order to be able to study the variation of effective exponents for this problem
in analogy with Figs.~\ref{fig:2},~\ref{fig:3}.

\begin{figure}
\centerline{\includegraphics[width=\figurewidth,clip=true]{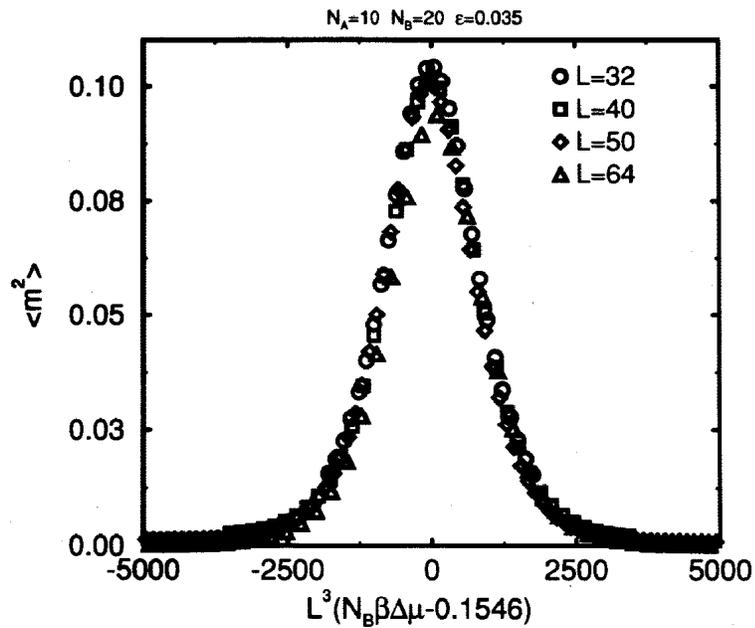}}
\caption{Finite-size scaling plot for the second moment of the order parameter
of an asymmetric polymer mixture ($N_A=10$, $N_B=20$) at a temperature
$T<T_{\rm c}$ ($\varepsilon=\varepsilon_{AB}/T=0.035$) as a function of the
normalized chemical potential difference, in order to locate $\Delta \mu_{\rm
coex}(T)$ by optimizing the ``data collapse'' for the range of values of $L$ as
indicated.  From Ref.~\protect\cite{20}.}
\label{fig:6}
\end{figure}

If one wishes to compare such simulations for polymer mixtures to experiments
on real systems~\cite{18,19}, an important complication that must be taken into
account is the asymmetry in chain length, $N_A\ne N_B$. This leads to two very
important technical complications: (i) While in the symmetrical case the
coexistence curve (including the critical point) occurs at a chemical potential
difference $\Delta\mu=0$, for $N_A\ne N_B$ phase coexistence occurs along a non
trivial curve $\Delta\mu_{\rm coex}(T)$ in the ($\Delta\mu,T$) plane, and hence
one has to search for the critical point ($\Delta\mu_{crit}=\Delta\mu_{\rm
coex}(T_{\rm c}),T_{\rm c}$) in a two dimensional variable space.
Fig.~\ref{fig:6}. shows that this problem can also be overcome by finite-size
scaling methods, utilizing the scaling behavior appropriate for first-order
transitions {$\Delta\mu-\Delta\mu_{\rm coex}(T)\propto L^{-d}$ in $d$
dimensions~\cite{48,49}} in order to locate $\Delta\mu_{\rm coex}(T)$\cite{20}.
(ii) Owing to the asymmetry, order parameter density and energy density become
coupled, and this ``field mixing'' effect needs to be disentangled from the
finite-size scaling analysis~\cite{21}. This problem is well known from
computer simulation of fluids and we shall not describe it here, but rather
draw attention to a recent review~\cite{67}.

\begin{figure}
\centerline{\includegraphics[width=\figurewidth,clip=true]{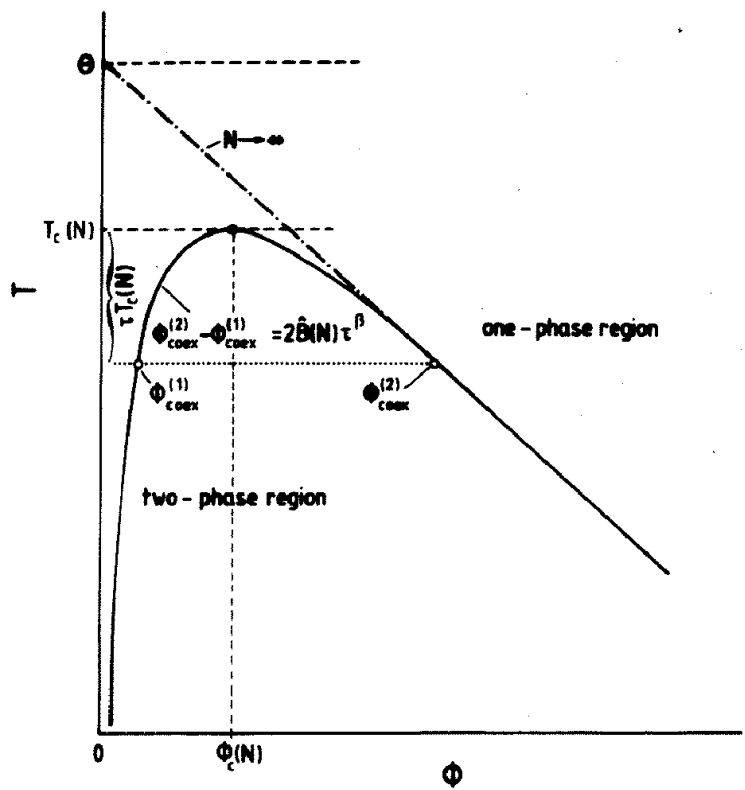}}
\caption{Schematic phase diagram of a polymer solution using the temperature
$T$ and the volume fraction $\phi$ taken by the effective monomers of the
polymer chains as variables. The coexistence curve separates a dilute solution
of collapsed chains (at $\phi_{\rm coex}^{(1)}$) from a semi-dilute solution of
overlapping chains (at $\phi_{\rm coex}^{(2)}$).  These two branches of the
coexistence curve merge at a critical point $T_{\rm c}(N)$, $\phi_{\rm c}(N)$.
For $N \to \infty$ this point merges with the $\Theta$-point of a polymer
solution at infinite dilution ($\phi \to 0$).}
\label{fig:7}
\end{figure}

\begin{figure}
\centerline{\includegraphics[width=\figurewidth]{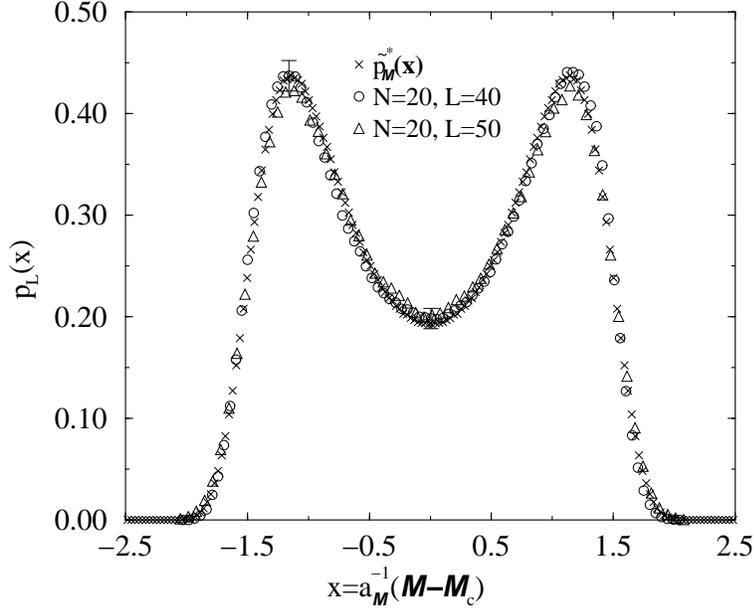}}
\caption{Critical order-parameter distribution for a polymer solution with
chain length $N=20$, modeled by the bond-fluctuation model on the simple-cubic
lattice, for linear dimensions $L=40$ and $L=50$, open symbols, and compared to
the order-parameter distribution of the Ising model (crosses). From
Ref.~\protect\cite{43}.}
\label{fig:8}
\end{figure}

This field-mixing problem is particularly severe for the unmixing of polymers
in solution beneath the $\Theta$-temperature (which formally can be considered
as a limiting case of a polymer mixture where $N_B=N, N_A=1$\cite{3}),
Fig.~\ref{fig:7}. However, by a suitable transformation of variables, one can
construct from $\phi$ and the energy density $u$, an appropriate field
$\mathcal{M}=(\phi-su)/(1-sr)$ (where $s,r$ are parameters that can be found
from a suitable analysis of the simulations, see~\cite{21,67}), which then
scales like the magnetization of the Ising ferromagnets. Figure~\ref{fig:8}
shows that the distributions of this variable at criticality nicely coincides
with the critical order parameter distribution of the Ising model (actually
this mapping can be used as a method for precisely locating the critical
point~\cite{21,67}).

From analyses of this kind it has been possible to obtain the critical
parameters of the model as a function of chain length, see e.g.\ 
Fig.~\ref{fig:8}. The simulation results reproduce nicely the behavior
$\rho_c\propto N^{-x}$ with $x\approx 0.37$ found also
experimentally~\cite{41,42}. However, the simulations also show that the chains
at the critical point are not yet partially collapsed, but are rather ideal,
and hence rule out the interpretation of this exponent value (which differs
from the classical results $x=1/2$~\cite{41}) as being due to the percolation
of partially collapsed chains. Consequently, the physical interpretation of
this exponent remains an open question~\cite{34,43}.

\section{Finite-size scaling above the upper critical dimension}

Remembering that the correlation length for $d>d^*=4$ has the mean-field
critical behavior $\xi_b = \xi_0 t^{-1/2}$, the free-energy density can be
written as~\cite{68}

\begin{equation}
f_L=
 L^{-d}\tilde{f}\left\{t\left(\frac{L}{\xi_0}\right)^2, 
                       uL^{4-d}, hL^{1+d/2}\right\} \;.
\end{equation}
Note that here exactly the same powers of $L$ appear as those for $l$ in
Eq.~(\ref{eq:enerscal1}). For $d>d^*$ there is only the Gaussian
fixed point to be considered. But although $u^*=0$ here and the power of
$L$ in the term $uL^{4-d}$ is negative so that $uL^{4-d}\to 0$ for
$L\to\infty$, the argument $uL^{4-d}$ must not be omitted: $u$ is a
``dangerous irrelevant variable''~\cite{69}, so when we consider the zero-field
susceptibility $\chi$ and the moment ratio $Q$ [Eq.~(\ref{eq:Q})], we find,
using $u \propto \ell_0^{d-4}$,

\begin{equation}
\chi=\left(\frac{\partial ^2 f_L}{\partial h^2}\right)_T=L^2
  P_\chi\left \{t\left
  ( \frac{L}{\xi_0}\right)^2,\left(\frac{L}{\ell_0}\right)^{4-d}\right\} \;,
\label{eq:pchi}
\end{equation}

\begin{equation}
Q=P_Q\left \{ t\left
    ( \frac{L}{\xi_0}\right)^2,\left(\frac{L}{\ell_0}\right)^{4-d}\right\} \;.
\label{eq:pq}
\end{equation}

Thus all scaling functions have two arguments, $t(L/\xi_0)^2$ and
$(L/\ell_0)^{4-d}$. However, it turns out~\cite{51} that a reduction to
one-variable scaling occurs for $L\to\infty$, namely

\begin{equation}
\chi \to \lim_{L\to\infty} 
  L^{d/2}\tilde{P}_\chi\left(tL^{d/2}\xi_0^{-2}\ell_0^{(4-d)/2}\right) \;,
\end{equation}

\begin{equation}
Q \to \lim_{L\to\infty}
  \tilde{P}_Q\left(tL^{d/2}\xi_0^{-2}\ell_0^{(4-d)/2}\right) =
  \tilde{P}_Q\{(L/\ell_T)^{d/2}\} \;,
\end{equation}
where in the last step we have introduced the ``thermodynamic length''
$\ell_T\propto t^{-2/d}$ \cite{52}, mentioned above.

Equations (13,14) can be understood from various arguments~\cite{51,52,53}.
Br\'ezin and Zinn-Justin argue~\cite{53} that in the initial Hamiltonian or the
corresponding statistical weight, one can treat the contribution from the
average magnetization $M$ separately,

\begin{equation}
\exp \left[-\mathcal{H} \{ s_i \}/k_BT \right] = 
  \exp\left\{-\frac{(M^2/M_b^2-1)^2}{8k_BT\chi_b/M_b^2 }L^d+\cdots \right\} \;,
\end{equation}
where the dots stand for contributions with non-uniform magnetization, i.e.\ 
fluctuations. Here $M_b,\chi_b$ are the mean-field bulk magnetization and
susceptibility, $M_b=\tilde{M_b}(-t)^{1/2}, \chi_b=\tilde{\chi_b}|t|^{-1}$. The
zero-mode theory neglects these fluctuations altogether and there the
distribution of the magnetization $P_L(M)$ scales as
\begin{equation}
P_L(M)\propto L^{d/2}\exp\{ -[M^2/(\tilde{M}_b^2(-t))-1]^2(L/\ell_T)^d/8 \} \;.
\end{equation}
From this result it is straightforward to derive the above scaling functions
$\tilde{P}_\chi$ and $\tilde{P}_Q$ explicitly~\cite{53}.

\begin{figure}
\centerline{\includegraphics[width=\figurewidth]{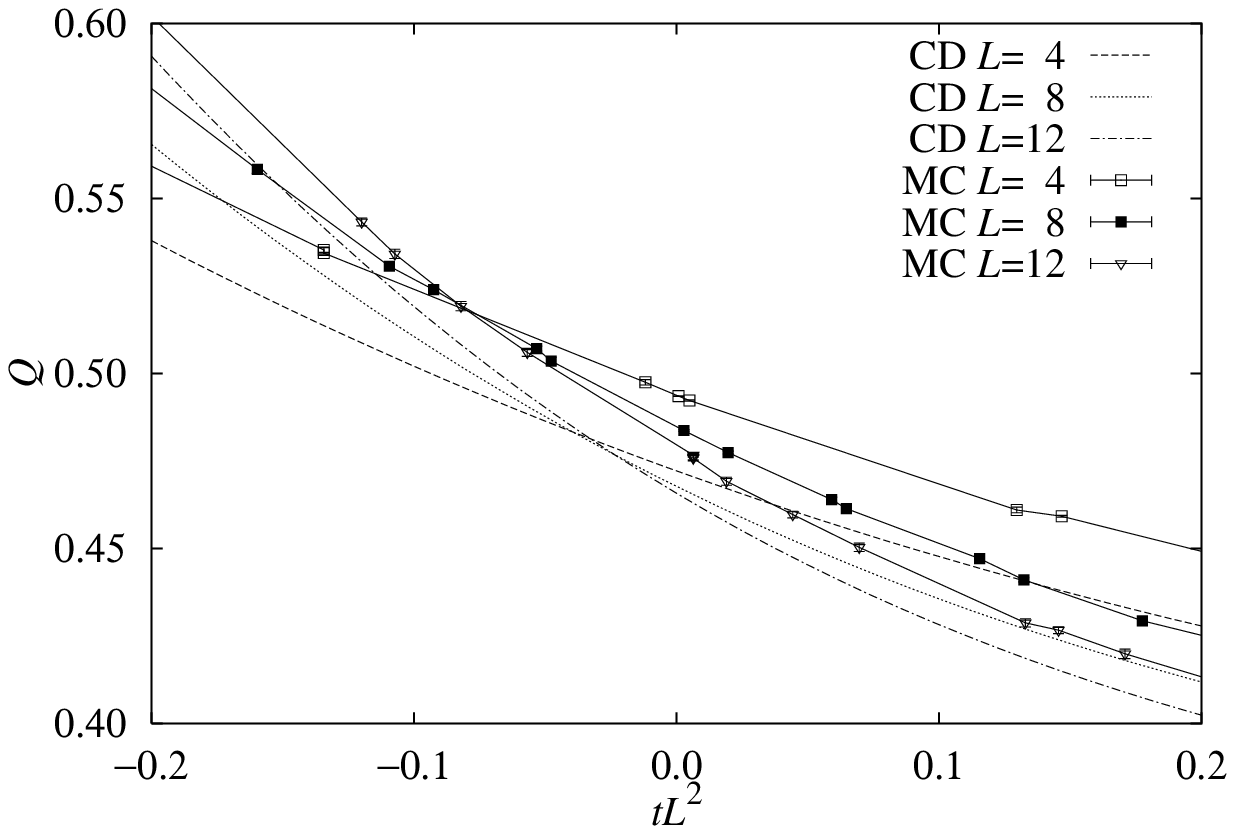}}
\centerline{\includegraphics[width=\figurewidth]{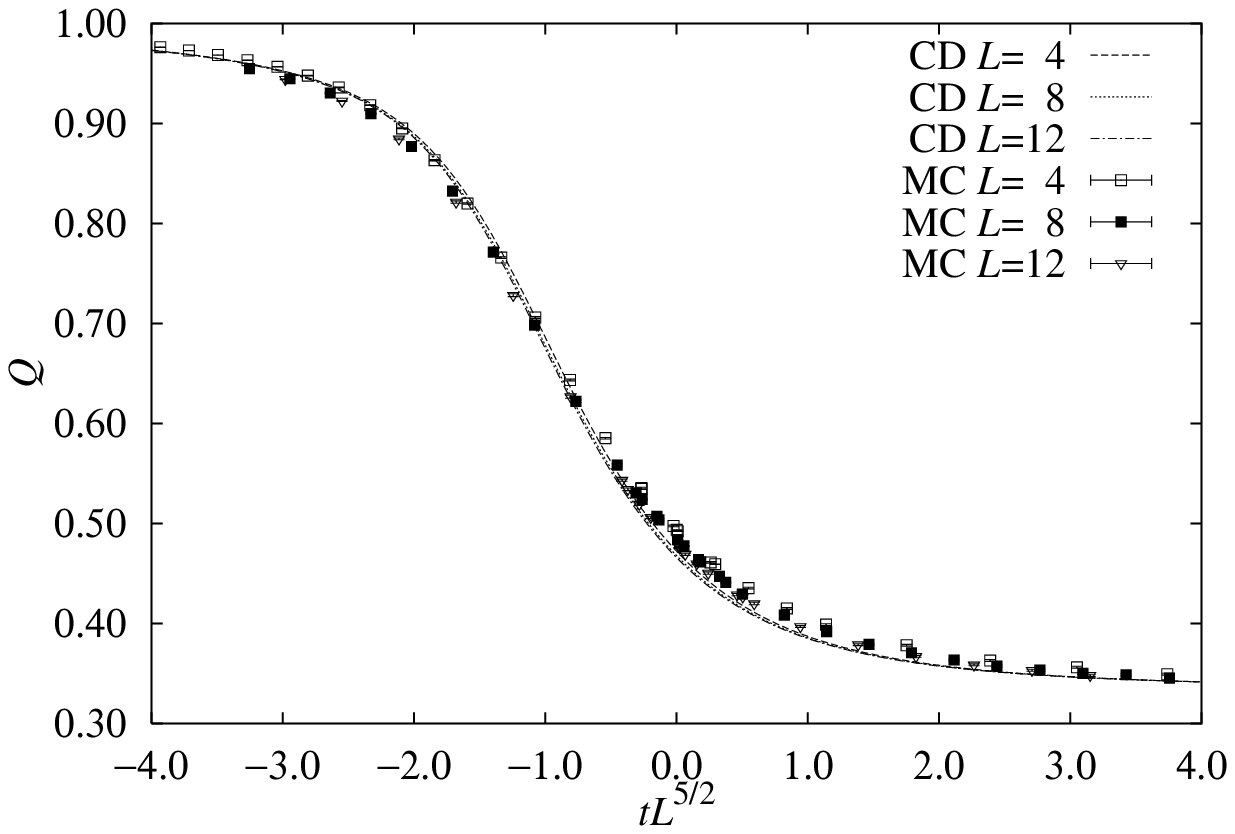}}
\caption{(a) Plot of $Q$ vs.\ $tL^2$ for the $d=5$ Ising model, demonstrating
the occurrence of spurious intersections both in the Monte Carlo
results~\protect\cite{61} and the Chen--Dohm theory~\protect\cite{60}.  (b)
Plot of $Q$ vs.\ the scaling variable $tL^{5/2}$; using parameters $\xi_0$,
$\ell_0$ extracted from various limits of the susceptibility~\protect\cite{61},
the Chen--Dohm theory can be evaluated without any adjustable parameter
whatsoever.  Note that for $L=12$ it is already graphically indistinguishable
from the ``zero-mode'' theory. From Ref.~\protect\cite{61}.}
\label{fig:10}
\end{figure}

Since this theory was proposed~\cite{51,52,53} it has
been a long-standing problem to verify the predictions by Monte Carlo
simulation. In particular, when one plots the moment ratio~$Q$ versus
temperature deviation from criticality, one should find a universal
intersection point at $T_{\rm c}$ at a value

\begin{equation}
\tilde{P}_Q(0)=8\pi^2/\Gamma^4(1/4)\simeq 0.456947 \;.
\end{equation}
However, the Monte Carlo results for small systems seem to intersect at a
different value $Q\simeq 0.52$ (Fig.~\ref{fig:10}). Also the temperature where
this intersection occurs is a little off, but since one does not know $T_{\rm
c}$ beforehand, one could simply imagine that the abscissa in Fig.~\ref{fig:10}
is mislabeled and $T_{\rm c}$ must be assigned differently.

\begin{figure}
\centerline{\includegraphics[width=\figurewidth]{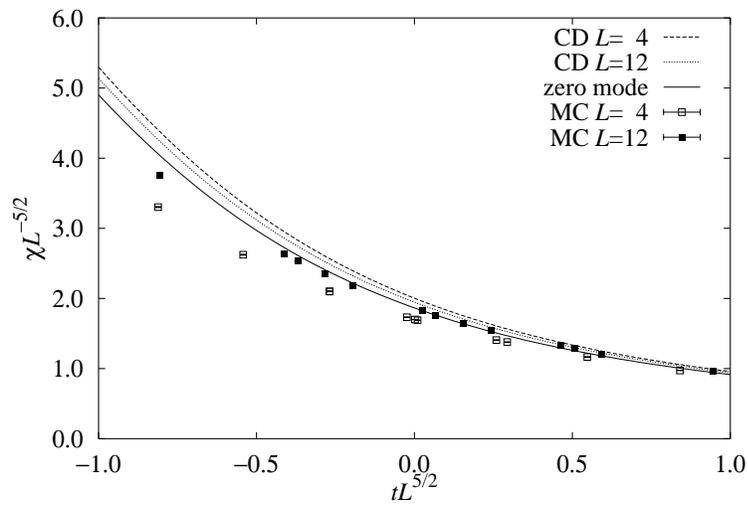}}
\caption{Plot of the scaled susceptibility $\chi L^{-5/2}$ vs.\ $tL^{5/2}$,
including the ``zero-mode'' result of Ref.~\protect\cite{53}, as well as the
predictions of Chen and Dohm~\protect\cite{60} evaluated for the same values of
$L$ as the Monte Carlo results shown. From Ref.~\protect\cite{61}.}
\label{fig:11}
\end{figure}

Chen and Dohm~\cite{59,60} have recently criticized the whole approach sketched
above and maintained that one must return to a finite-size scaling
description in which both variables $t(L/\xi)^2$ and $(L/\ell_0)^{4-d}$ are
kept separate, as in Eqs.\ (\ref{eq:pchi}) and~(\ref{eq:pq}).  They also
obtained the scaling functions $P_Q$ and~$P_\chi$ in a first-order loop
expansion as a function of these variables.  Indeed their result is
qualitatively similar to the Monte Carlo data (Fig.~\ref{fig:10}, broken
curves), although in quantitative respects their treatment offers little
improvement. This is seen, for instance, in a scaling plot of the
susceptibility: The Chen--Dohm theory approaches the zero-mode results from
above, while in the regime of interest the Monte Carlo data fall below the
zero-mode result (Fig.~\ref{fig:11}). These discrepancies remain present for
considerably larger $L$ than shown here~\cite{61}.

Thus we arrive at a rather disappointing state of affairs---although for the
$\phi^4$ theory in $d=5$ dimensions all exponents are known, including those of
the corrections to scaling, and in principle very complete analytical
calculations are possible, the existing theories clearly are not so good.
Perhaps the discrepancies result because the theory of Ref.~\cite{60} is only
one-loop order, perhaps because other corrections are missing. While presumably
the zero-mode one-parameter scaling is true asymptotically for $L \to \infty$,
the corrections to this limit disappear only rather slowly, as
Fig.~\ref{fig:10}(a) has demonstrated.

\section{Concluding remarks}

\begin{figure}
\centerline{\includegraphics[width=\figurewidth]{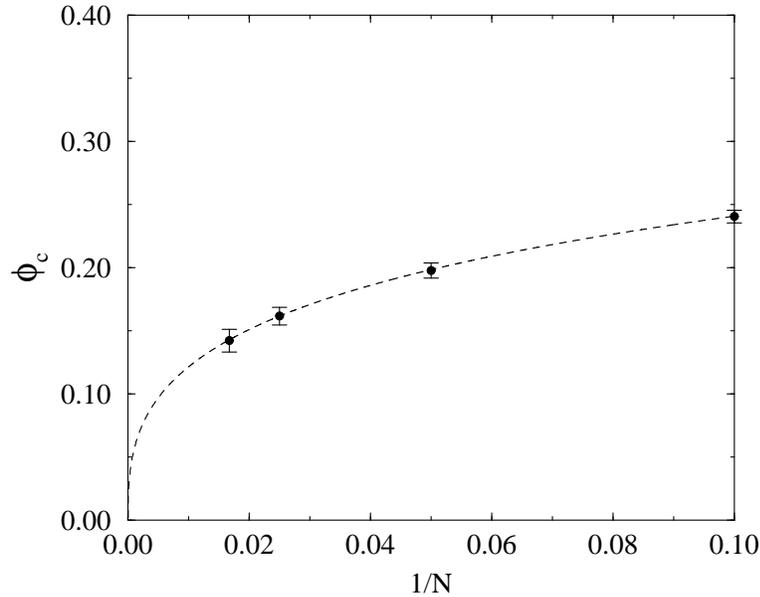}}
\caption{Estimates of the critical volume fraction $\phi_{\rm c}$ of monomers
for a polymer solution (modeled by the bond-fluctuation model on the
simple-cubic lattice, with an attractive energy between monomers at distances
$r \leq \sqrt{6}$) as a function of the inverse chain length.  The broken
curven represents a fit of the form $\phi_{\rm c}=(1.1126+1.3N^{0.369})^{-1}$.
From Ref.~\protect\cite{43}.}
\label{fig:9}
\end{figure}

While the estimates of the critical exponents for the $d=3$ Ising model are
impressively accurate~\cite{70,71,72} and analytical~\cite{70} and Monte Carlo
\cite{71,72} estimates agree within very small error margins, the situation is
different for the problems considered in the present paper: Analytical work is
restricted to low-order $\varepsilon$-expansions or low-order loop-expansions
and discrepancies between theory and simulation occur that are not fully
understood. More work will be needed to clarify the situation. Note that the
Ising to mean-field crossover considered here really is the simplest example of
crossover phenomena, since the crossover exponent is rigorously
known---crossover from one nontrivial fixed point to another is presumably more
tricky to deal with. And for problems such as the critical point of polymer
solutions, even the proper theoretical approach is controversial, and hence it
is unclear whether the exponent $x \approx 0.37$ (Fig.~\ref{fig:9}) is a
universal property at all~\cite{34,35,36,37,38,39,40,41,42,43}.

Further problems appear when one is not concerned with bulk critical phenomena
in ideal, homogeneous systems, but when one considers inhomogeneous systems,
e.g., systems with random quenched disorder (e.g., Ising and Potts models
exposed to random fields, spin glasses, etc.~\cite{73}). For instance, for a
Potts spin glass finite-size scaling is not even understood on the mean-field
level, at least for cases where first-order transitions without latent heat
occur~\cite{74}. Also for systems with a regular inhomogeneity, e.g., Ising
films with competing walls which allow for interface
localization--delocalization transitions, one has fascinating critical behavior
and crossover, of which the details still need to be unraveled~\cite{75}.  Thus
the Monte Carlo investigation of phase transitions---both in equilibrium and in
driven systems~\cite{76}---will remain an active and challenging field.

\subsection*{Acknowledgments}
\begin{figure}
\centerline{\includegraphics[width=\figurewidth,clip=true]{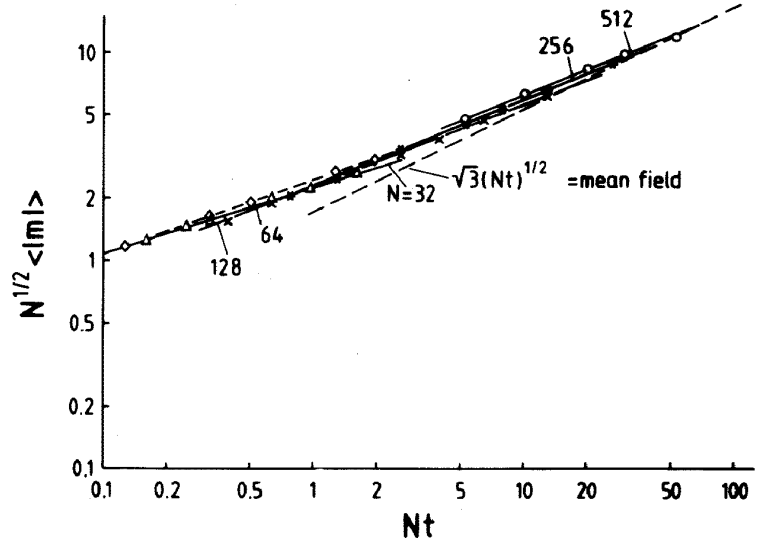}}
\caption{Crossover scaling plot for the order parameter $\langle |m|\rangle =
\langle|\phi_A-\phi_B|\rangle / (\phi_A+\phi_B)$ of a binary polymer
mixture~(A,B) with symmetrical chain lengths $N_A=N_B=N$.  $\phi_A, \phi_B$ are
the volume fractions of $A$ and $B$ monomers, respectively. The points are
simulation results for the bond-fluctuation model on a simple-cubic lattice,
using concentration $\phi_v=0.5$ of vacant sites. Straight lines in this
log--log plot indicate power laws with effective exponents, $\langle m \rangle
= \hat{B}_{\rm eff} t^{\beta_{\rm eff}}$, $t=1-T/T_{\rm c}$. The broken
straight line shows the mean-field result, $\langle m \rangle
=\sqrt{3}t^{1/2}$, to which the data converge for $N \to \infty$. From
Ref.~\protect\cite{17}.}
\label{fig:5}
\end{figure}

We thank the HLRZ J\"ulich for access to a Cray-T3E where part of the
computations have been performed.  One of us (K.B.) thanks H.-P. Deutsch for a
fruitful collaboration that led to the results shown in Fig.~\ref{fig:5}.

\newpage

\end{document}